\documentstyle[12pt]{article}
\begin{document}
\baselineskip 1.5 \baselineskip
\vspace{1cm}
\begin{center}
     {\Large Composite Operators and Topological Contributions in Gauge Theory }
\end{center}
 
\vspace{1cm}
 
\begin{center}
     Jungjai  Lee \footnote{E-mail : jjlee@road.daejin.ac.kr} \\
 
\vspace{0.3cm}
     {\em  Department of Physics, DaeJin University, PoCheon, GyeongGi 487-711, Korea}
\end{center}
 
\vspace{0.8cm}
 
\begin{center}
     Yeong Deok  Han \footnote{ E-mail : ydhan@core.woosuk.ac.kr} \\
\vspace{0.3cm}
     {\em Department of Physics, Woosuk University,
     Hujeong, Samrye, Wanju, Cheonbuk 565-701, Korea }
\end{center}
 
\vspace{0.8cm}
 
\begin{flushleft}
{\bf Abstract }
\end{flushleft}
In $D$-dimensional gauge theory with a kinetic term based on the p-form tensor gauge field,
we introduce a gauge invariant operator associated with the composite formed from
a electric $(p-1)$-brane and a magnetic $(q-1)$-brane in $D=p+q+1$ spacetime dimensions.
By evaluating the partition function for this operator, we show that the expectation value of this operator
gives rise to the topological contributions identical to those in
gauge theory with a topological Chern-Simons BF term.

\newpage
 
 The exotic statistics, anyons and fermion-boson transmutation for the
composite state of a point charge and a magnetic vortex has been discussed by Wilczek\cite{a}.
In $(2+1)$ dimensional spacetime, the anyons have a well-known
physical realization where magnetic flux tubes are attached to
charged particles and the Aharonov-Bohm phase resulting from
adiabatic transport of the composites give them fractional
exchange statistics. It was generalized to the fact that the composite of a closed Nambu charged
string and a point vortex presented the unusal statistics in (3+1)
dimensions\cite{b}.
 
 Also, Polyakov showed the fermion-boson transmutation in (2+1) dimensions
by investigating the small momentum behavior of a scalar field interacting with
topological Chern-simons term\cite{c}.
This Chern-Simons mechanism of statistical transmutation was shown
to hold for string-like object interacting with topological BF term in (3+1) dimensions\cite{d} \cite{e}.
The topological quantum field theory which gives the appropriate
generalization to higher dimensions has been studied\cite{f}.
Chern-Simons theory gives the representations of linking numbers
of curves in $3$ dimensions, BF theories provide the path integral
representations of the linking and intersection numbers of generic
surfaces in D dimensions\cite{e} \cite{f}.
 
 In these statistical phenomena like fermion-boson transmutation, even
though the topological term(Chern-Simons term or BF term in higher
dimensions) plays an essential role, such topological contributions
may arise from the gauge theory with the kinetic term as electromagnetism\cite{g}.
In Wilczek's work\cite{a}, if two composites of a charged particle and
a point vortex are interchanged, an additional phase factor appears
in all gauge invariant observables, this is because each composite should
be covariantly transported in the gauge potential of the other. So the composites have the unusual statistics.
 
Recently there has increasingly been the theoretical evidence for the extended
objects( string, membrane,..., $p$-brane etc.) as the fundamental
constituents of the universe\cite{h}. In this paper, in order to see if the topological
contributions arise from the higher dimensional gauge theories with the minimal Maxwell kinetic term,
we will apply Polyakov's path integral arguments\cite{c} to the composite
of electric $(p-1)$-brane and magnetic $(q-1)$-brane in D(=p+q+1) spacetime dimensions.
The application to (2+1) or (3+1) dimensional gauge theory
showed that the expectation value of the composite operators involves some topological contributions
associated with link invariants \cite{g} \cite{i}.

We start by introducing an gauge invariant operator associated with the composite of
an electric $(p-1)$-brane and a magnetic $(q-1)$-brane in D space-time dimensions
 
\begin{equation}
  X(W_q ; W_p) =  exp ( -i \int_{W_q}*(F_{p+1})) ~ exp(i \int_{W_p} A_p ).
\end{equation}
This operator depend on the $p$-dimensional world volume $W_p$
and $q$-dimensional world volume $W_q$ in $D$ spacetime dimensions.
Here $A_p$ is a $p$-form gauge field which couples to electric $(p-1)$-brane, and
$F_{p+1}$ is the corresponding $(p+1)$-form field strength.
The operator $X(W_q ; W_p)$ should be called the extended 't Hooft-Wilson operator.
In fact, the first factor is the covariant version of the 't Hooft operator
originated from the closed hyper-path of $(q-1)$-brane, thus it should be satisfied that $D-p-1=q$,
because $*(F_{p+1})$ is $(D-p-1)$-form field in $D$ dimensions.
 
Our aim is to calculate the transition amplitude for the pair of a
$(p-1)$-brane and a $(q-1)$-brane, and to investigate the form of
the partition function, and then to see how the topological contributions arise.
When the $(p-1)$-brane and $(q-1)$-brane of the composite evolve
from the initial configuration $P_i$ to the final configuration
$P_f$, we have the transition amplitude
 
\begin{equation}
  G(P_i , P_f )= \sum_{ W_p (P_i , P_f) W_q (P_i , P_f )} e^{-S_p - S_q } < X_E ( W_p , W_q) >
,
\end{equation}
where the sum is over all $p$-dimensional world volume $W_p$ and
$q$-dimensional world volume $W_q$ interpolating between the
initial and the final configurations $P_i$ and $P_f$ of the $(p-1,q-1)$-brane composite.
The purely geometrical action $S_p$ for a $(p-1)$-brane with a classical trajectory, parametrized  by
 
\begin{equation}
  x^{\mu} = x^{\mu} ( \xi^{0}, ... \xi_{p-1} )
,
\end{equation}
is the $p$-dimension world-volume induced on the
trajectory by the external $D$ dimensional metric times
$(p-1)$-brane tension $T_p$
\begin{equation}
  S_p  = T_p \int_{W_p} d(Vol)
,
\end{equation}
where the Riemannian volume element is given in terms of the
determinant of the world-volume metric $h = det(h_{ij})$, by
\begin{equation}
  d(Vol) = d\xi^{p-1} \wedge ... \wedge d\xi^{0} \sqrt{|h|}
.
 \end{equation}
And $<X_E (W_p , W_q)>$ is the euclidean form of vacuum expectation value of the
operator $X$ by the minimal "Maxwell" action for Abelian $p$-forms
\begin{equation}
  S_A =\int_{V_D} d^D x (*F)_{D-p-1} \wedge F_{p+1}
.
\end{equation}
 
The partition function is given as follows
\begin{equation}
  Z = \sum_{ W_p ,~ W_q} e^{-S_p - S_q} < X_E ( W_p , W_q) >
\end{equation}
with
\begin{eqnarray}
  <X_E (W_p , W_q )> &=& \int DA_p ~ exp( - {1 \over 2(p+1)! }
  \int d^D x F_{p+1}^2 ) \nonumber \\
  &\times& exp(-i \int_{W_q } *F ) ~ exp ( i \int_{W_p } A_p )
,
\end{eqnarray}
where $S_q$ is the geometrical action for a ($q-1$)-brane.
 
Performing the integration over $A_{\mu_1 \mu_2 ... \mu_p }$ leads to
\begin{eqnarray}
&<&X_E (W_p , W_q )> = exp \left( \int_{W_p} dx^{\mu_1 \mu_2 ... \mu_p } \int_{W_p} dy_{\mu_1 \mu_2 ... \mu_p }
 {1 \over {|x-y|^{D-2}}}\right) \nonumber \\
 &\times& exp \left(i \int_{W_q} dx_{\mu_{p+2} ... \mu_D } \int_{W_p} dy^{\mu_2 ... \mu_{p+1} }
 \epsilon^{\mu_1 \mu_2 ... \mu_{p+1} \mu_{p+2} ...\mu_D } {\partial \over {\partial x^{\mu_1}}}
 ({1 \over {|x-y|^{D-2}}}) \right) \nonumber \\
 &\times& exp \left(\int_{W_q} dx^{\mu_{p+2} ... \mu_D } \int_{W_q} dy_{\mu_{p+2} ... \mu_D} ~ \delta^{(D)} (x-y) \right)
,
\end{eqnarray}
where $G(x,y)$ is a Green's function of Laplacian operator in D
euclidean dimensions and its singular structure is given as that $G(x, y) \sim |x-y|^{-(D-2)}$.
 
The second term in the exponetial is equal to the linking
number between the $p$-dimensional object and the $q$-dimensional
object in $D=p+q+1$ dimensional spacetime. we can understand it from
the theorem given by the intersection number\cite{f}. If $\partial \Sigma $
and $\partial \Sigma'$ are disjoint compact and oriented
$p$- and $(D-p-1)$-dimensional boundaries of two oriented submanifolds
of an D-dimensional oriented manifold $M^D$, the linking number of
$\partial \Sigma $ and $\partial \Sigma'$ is able to be defined
as the intersection number of $\Sigma $ and $\partial \Sigma'$
or $\partial \Sigma $ and $ \Sigma'$. As the dimension of
(say) $\Sigma $ is equal to the codimension of $\partial
\Sigma'$, they will generically intersect transversally at isolated
point $x_i$. The intersection number of $\Sigma $ and $\partial \Sigma'$
is then defined as ${\rm INT} (\Sigma , \partial \Sigma') = \sum_i sign (x_i)$
which we also identify with the linking number ${\rm LINK}(\partial \Sigma , \partial \Sigma')$[6].
 
The other two terms in the exponential require a regularization.
Following the regularization procedure by using "point splitting"
these contribute to the renormalization of $p$-brane or $q$-brane tentsions.
To do this regularization, we describe the extrinsic geometry of
the $p$-brane. At a point $\xi$ on the $(p+1)$-dimensional
hyper-surface, we define the tangent plane to be spanned by the
vectors ${\bf e}_a (\xi)$, $a=1,2,...,p+1$. The vectors normal to
the tangent plane are denoted by ${\bf n}_i (\xi )$, $i=1,2,..., D-p-1$.
For each hyper-surface $W_p$ parametrized by $x^{\mu} (\xi)$, we
introduce a framing surface parametized by $y^{\mu} = x^{\mu} (\xi) + \epsilon \hat{n}^{\mu} (\xi)$
$(\epsilon > 0, | \hat{\bf n} (\xi)|^2 = 1)$, where $\hat{n}^{\mu}$ is
a vector field normal to the hyper-surface $W_p$. This
regularization contributes to the renormalization of the string $T_p \rightarrow T_p^R$.
 
In discussion, the classical action of this gauge theory is not
topological, how does it give rise to the topological contributions
identical to those of topological gauge theory with a Chern-Simons
term (or BF term in higher dimensions)? Though the action is
dynamical, the operator being considered is the composite formed
from a electric $(p-1)$-brane and a magnetic $(q-1)$-brane to which a $p$-form gauge field couples
in $D$ dimensions. So, when they evolve and form a closed hyper-path,
they may be mutually linked. That is, if two composites are interchanged,
each composite should be covariantly transported in the $p$-form gauge potential of the other.
Thus the topological contributions are originated from the electro-magnetic dual
structure of the composite source. In more mathematical details,
the propagator in the topological Chern-Simons theory is given as
the inverse operator of the differential operator "$\epsilon
\partial$", that is the propagator contains the anti symmetric $\epsilon$
tensor, but the gauge invariant source term does not. On the other
hand, in the present gauge theory with a kinetic term,
the propagator which is the inverse of Laplacian operator does not contain  $\epsilon$ tensor,
while one of the current sources  $J$ and $K$ to which the gauge field couples does
it. If we write the ineraction terms in details,
\begin{eqnarray}
    S_I &=& -i\int_{W_q} (*F)_{D-p-1} + i \int_{W_p} A_{p} \nonumber \\
        &=& \int d^D x (J + K)^{\mu_1 \mu_2 ... \mu_p }A_{\mu_1 \mu_2 ... \mu_p }
,
\end{eqnarray}
where
\begin{equation}
    J^{\mu_2 ... \mu_{p+1} } (x)= {{-i} \over {(D-p-1)!}} \int_{W_q} \epsilon^{\mu_1 \mu_2 ... \mu_{p+1} \mu_{p+2}...\mu_D}
    {\partial \over {\partial y^{\mu_1}}} \delta^{(D)} (x-y)
    dy^{\mu_{p+2}...\mu_D}
,
\end{equation}
 
\begin{equation}
    K^{\mu_1 ... \mu_p }(x) = i \int_{W_p} \delta (x-y) dy^{\mu_1 ... \mu_p}
.
\end{equation}
Therefore the expectation value of composite operator can give
the linking number identical to that in topological BF theory.
 
In this theory, since the linking number is not self-linking one, in any space time dimensions
these topological phases can be occurred. In higher dimensional Chern-Simons theory
the possible topological contributions steming from the self linking of the
single gauge invariant operator, however, are limited only to the spacetime dimensions $D=3, 7, 11, 15$,
since Chern-Simons term can be constructed only in odd dimensions and
the self-linking of the object with even dimensional structure vanish\cite{j} \cite{k}.
 
In conclusions,
we have shown the topological contributions from the expectation
values of the composite operator $X(W_q , W_p)$ in the gauge
theory with a minimal kinetic term  are identical to those obtained
for Wilson hypersurfaces in topological quantum field theory with
a BF Chern-Simons term in D dimensions. This topological phase
factors originate from the distinct ways in which the $(p-1)$-brane
world surface may braid around the $(q-1)$-brane world surface when they evolve
in $D$ dimensional space-time. This result can be generalized to the
non-intersecting multi $p$-brane hypersurfaces and the non-Abelian extension deserves futher
investigations. Since statistics has barely been touched for
objects such as string, membrane, p-brane, it is very tempting to
analyze the generalized statistics in different contexts.

\section*{Acknowledgements}
 
This work was supported by the Research Fund of Woosuk University.

\end{document}